\documentclass[12pt,a4paper]{article}
\usepackage{latexsym,amssymb,amsmath}

\begin{document}
\title{A note on the fast power series' exponential}
\date{}
\author{I. S. Sergeev}

\maketitle
\begin{abstract}
It is shown that the exponential of a complex power series up to
order $n$ can be implemented via $(23/12+o(1))M(n)$ binary
arithmetic operations over $\mathbb C$, where $M(n)$ stands for
the (smoothed) complexity of multiplication of polynomials of
degree $<n$ in FFT-model. Yet, it is shown how to raise a power
series to a constant power with the complexity $(27/8+o(1))M(n)$.
\end{abstract}

\section{Introduction}

It is very well known that the exponential of a power series (as
well as some other elementary operations) has the same order of
complexity as multiplication, see e.g.~\cite{br,gg}. (When we
speak about complexity we consider a computational model of
circuits or straight-line programs over arithmetic basis $\{ \pm,
* \} \cup \{ ax \mid a \in \mathbb C \}$, see e.g.~\cite{gg}. The field should not be
necessary complex, it might be real or any algebraically closed,
or any other which supports appropriate FFT.)

Previous papers set up a convention to estimate complexity of the
basic power series' operations (including exponential) in the
number of multiplications of the same size. The very last papers
(since 2000) use the special FFT-way multiplications.

The complexity function $M(n)$ of FFT-way multiplication can be
introduced as follows. Let $F^{\circ}(n)$ be the complexity of FFT
of order $n$ over $\mathbb C$. Let $F(n)$ be its smoothed version,
that is, $F(n) = \min_{m \ge n} F^{\circ}(m)$. For simplicity we
also assume $F(n)=\Omega(n\log n)$ (in fact, $F(n)=\omega(n)$ is
sufficient) and $kF(n)=(1+o(1))F(kn)$ for $1 \le k \le \log
n$.\footnote{These assumptions are for more convenient way of
writing final complexity bounds only, it does not affect any other
aspect of the proof. Notation $g=\omega(f)$ means $f=o(g)$.} Then,
let $M(n) = 6F(n)$. In any case $M(n)$ serves as an upper
asymptotic estimate of the complexity of multiplication of
polynomials of degree $<n$.

Write an upper estimate of the complexity of computing exponent of
a power series in $\mathbb C[[x]]$ modulo $x^n$ in the form
$(A+o(1))M(n)$. Let us list some previously known results:
$A=17/6$~\cite{b2}, $A=14/3$~\cite{vdh,vdh2},
$A=13/6$~\cite{iss,is10} and~\cite{h}. Next, we will show that
$A=23/12$ is also valid. The method is a straightforward
combination of methods~\cite{iss,is10} and~\cite{h}. To be more
precise, scheme of computation follows~\cite{h} and technique is
inherited from~\cite{iss,is10}.

We will also show that one can compute a constant power of a power
series in $\mathbb C[[x]]$ modulo $x^n$ with the complexity
$(B+o(1))M(n)$, where $B=27/8$. It slightly improves the
previously best known factor $B=41/12$~\cite{iss,is10}.

Some notation. Let $f \in \mathbb C[[x]]$, then $f_{..n}$ denotes
$f \bmod x^n$ and $\lfloor f/x^n \rfloor$ stands for
$(f-f_{..n})/x^n$. If $f = \sum_{i \ge 0} f_i x^i$, then $\Delta
f$, ${\bf J} f$, $\ln f$ (when $f_0=1$) and $e^f$ (when $f_0=0$)
denotes formal derivative, formal integral, formal logarithm and
formal exponent respectively:
$$ \Delta f = \sum_{i \ge 1} if_i x^{i-1}, \quad {\bf J} f = \sum_{i \ge 1} \frac{f_{i-1}}i x^{i},
\quad  \ln f = - \sum_{i \ge 1} \frac{(1-f)^i}{i},  \quad e^f =
\sum_{i \ge 0} \frac{f^i}{i!}. $$

\section{Exponent}

Consider a problem of computing exponent of a power series $h$,
$h_{..1}=0$. Denote $f=e^h$, $r=1/f$. Recall that $\Delta h=\Delta
f/f$.

The next iterative formula~\cite{vdh} (derived as a solution of an
equation $\Delta f = g f$ with $g=\Delta h$ in this case) is valid
for $m\ge n$:
\begin{equation}
f_{..m+n} = f_{..m} + f_{..n} \, {\bf J} \left( x^{m-1}r_{..n}
\left\lfloor \Delta (h_{..m+n}) f_{..m} / x^{m-1} \right\rfloor
\right) \mod x^{m+n}.
\end{equation}

Let $E(n)$ and $I(n)$ denote the complexity of computation $e^f$
and $1/f$ modulo $x^n$ respectively. Then we can use~(1) to
compute $f_{..m}$ with the complexity
\begin{equation}
E(n) + I(n) + (13+o(1))F(m) \sim (13/6+o(1))M(m)
\end{equation}
for appropriately chosen parameters $m$ and $n$, e.g. $n=o(m)$ and
$m=O(n\sqrt{\log n})$. (That is one of the ways to obtain factor
$13/6$ in the complexity estimate for exponent~\cite{is10}.)

To achieve the complexity bound~(2) split series into blocks of
appropriate size $k$, e.g. $k \in o(n) \cap \Omega(n/\sqrt{\log
n})$. Then use double DFT of order $(2k,k)$ to perform block
multiplications in~(1).

{\it Double DFT} of order $(l,k)$ as a map from $\mathbb C[x]$ to
$\mathbb C^{l+k}$ is defined so that its first $l$ components are
the components of DFT of order $l$, another $k$ components are the
components of composition of the variable substitution $x \to
\zeta x$ and DFT of order $k$, where $\zeta$~--- an appropriate
complex number. Multiple DFT can be defined in similar
way~\cite{is9}. Multiple DFTs are useful to perform
multiplications of different sizes on the overlapping sets of
inputs. Double DFT of order $(l,k)$ or its inverse costs as much
as DFT of order $l$ and DFT of order $k$ plus $O(l+k)$ extra
operations, see~\cite{is10} for details.

In the case $l=2k$ (as we have) one can use an ordinary DFT of
order $3k$ decomposed into outer DFTs of order $3$ and inner DFTs
of order $k$ instead.

The main term of the complexity estimate~(2) is contributed by:
\linebreak $3(m/k)F(k)$ (which we assume to be approximately
$3F(m)$) operations to compute DFTs of blocks of $f$, the same
number of operations to compute DFTs of blocks of $\Delta h$, the
same number of operations to restore blocks of the triple product
under the integral, $2F(m)$ operations to compute $2k$-order parts
of DFTs of blocks of ${\bf J}(\ldots)$ and the same number of
operations to restore the product $f_{..n} \, {\bf J} (\ldots)$.
Other steps (precomputation of $f_{..n}$ and $r_{..n}$, additions,
implementation of $\Delta$ and $\bf J$ operators, calculations in
the DFT-image spaces) contribute $o(m\log m)$ in total complexity.

To provide a hint for verification we consider a subproblem of the
triple product computation (this step seems to be less evident in
the algorithm above) in Appendix. Other details (if necessary) see
in~\cite{is10}.

Next we turn to introduce an improved version of the algorithm
with the use of idea due to D.~Harvey~\cite{h}.

Suppose we are to compute $f_{..2m}$. Firstly we compute $f_{..m}$
acting as mentioned above. By the way we also have DFTs of blocks
of $f_{..m-n}$ and $\Delta (h_{..m})$ been computed. At the second
stage we use formula~\cite{br}
\begin{equation}
f_{..2m} = f_{..m} + f_{..m}(h - \ln f_{..m})_{..2m} \mod x^{2m}
\end{equation}
derived by the discrete Newton---Raphson method as the solution of
an equation $\varphi[f]=h$ with $\varphi[x]= \ln x$ in our case.
This stage generally consists of the two essential parts:
computation of $\ln f_{..m}$ up to order $2m$ and final
multiplication $f_{..m}$ by $h-\ln f_{..m}$.

Denote $s=\Delta(f_{..m})/f_{..m}$. We compute $(\ln
f_{..m})_{..2m}$ as ${\bf J} s_{..2m-1}$ via the
iteration~\cite{km}
\begin{equation}
 s_{..m'+n-1} = s_{..m'-1} - x^{m'-1}r_{..n}\left\lfloor s_{..m'-1}f_{..m} / x^{m'-1} \right\rfloor \mod x^{m'+n-1},
\end{equation}
where $m'\ge m$. We start from $s_{..m-1} = \Delta (h_{..m})$.

To perform the calculations by~(4) we use $(6+o(1))F(m)$ extra
operations: half of them to compute DFTs of remaining blocks of
$s$, another half to compute blocks of the triple product (see
Appendix for some details).

To complete the computation of $f_{..2m}$ we use another
$(4+o(1))F(m)$ operations: half of them to compute $2k$-order DFTs
of the order of blocks of $h-\ln f_{..m}$ and roughly the same
number to restore blocks of the product in~(3). Recall that
$2k$-order DFTs of almost all blocks of $f_{..m}$ are also
computed at the first stage of the algorithm since we use double
DFTs (this is the only place we gain a benefit from exploiting
double DFT).

To summarize, we can compute the exponent up to order $2m$ with
the complexity $(23+o(1))F(m)$.

\section{Exponentiation}

Consider a problem of raising of a power series $h$, $h_{..1}=1$
to a power $C \in \mathbb C$. Denote $f=h^C$, $r=1/f$, $\rho =
1/h$, $s=C\Delta h/h$.

The way of computing a power is just similar with that of
computing an exponent. We will give only a sketch.

To compute the first half of the required coefficients of $f$ we
use the formula
\begin{equation}
f_{..m+n} = f_{..m} + f_{..n} \, {\bf J} \left( x^{m-1} r_{..n}
\left\lfloor s_{..m+n-1}f_{..m} / x^{m-1} \right\rfloor \right)
\mod x^{m+n}
\end{equation}
derived as a solution of the equation $\Delta f = s f$. Next, we
switch to the formula
\begin{equation}
f_{..2m} = f_{..m} + f_{..m}({\bf J} s - \ln f_{..m})_{..2m} \mod
x^{2m}
\end{equation}
derived as a solution of the equation $\ln f = C\ln h$.

To solve a subproblem of computing $s$ we use the
iteration~\cite{km}
\begin{equation}
 s_{..m+n-1} = s_{..m-1} + \rho_{..n}(\Delta(h_{..m+n}) - s_{..m-1}h_{..m+n}) \bmod x^{m+n}.
\end{equation}

As above all series are divided into blocks of size $k$, except
that the first half of the series $\Delta h$ is divided into
blocks of size $2k$. We also use double DFTs of order $(2k,k)$.

Suppose we are given $f_{..n}$, $r_{..n}$, $\rho_{..n}$ and
$s_{..n-1}$. Then we can compute $f_{..2m}$ using
$(40.5+o(1))F(m)$ operations. This bound is contributed by the
following parts:

$(10.5+o(1))F(m)$ operations to compute $s_{..m-1}$ and DFTs of
blocks of $s_{..m-1}$ via~(7). We use DFTs of blocks of $h$, $s$,
$\Delta h$, $\rho$ and inverse DFTs to restore triple
multiplications (it is essential that the blocks of
$\Delta(h_{..m})$ are double-sized);

$(10+o(1))F(m)$ operations to compute another half of
$s_{..2m-1}$. Here each iteration~(7) is performed via two
ordinary multiplications with the use of DFTs of order $2k$;

$(10+o(1))F(m)$ operations to compute $f_{..m}$ and DTFs of its
blocks by~(5). The procedure is the same as in the first part of
the exponential algorithm;

$(6+o(1))F(m)$ operations to compute $\ln f_{..m}$ up to order
$2m$. This step coincides with that of the exponential algorithm
above;

$(4+o(1))F(m)$ operations to perform final multiplication in~(6)
via DFTs of order $2k$.

Therefore, we have got the required complexity estimate.

Research supported in part by RFBR, grants 11--01--00508,
11--01--00792, and OMN RAS ``Algebraic and combinatorial methods
of mathematical cybernetics and information systems of new
generation'' program (project \linebreak ``Problems of optimal
synthesis of control systems'').

\section*{Appendix}

Let $f,g,h \in \mathbb C[[x]]$. Consider a problem of computing
$$ q = f \lfloor g h / x^m \rfloor \mod x^n, $$
where $m$ is a multiple of $n$. Suppose series $f,g,h \in \mathbb
C[[x]]$ are given divided into blocks of size $k$ (we assume for
simplicity that $n$ is a multiple of $k$):
$$  f = \sum_{i \ge 0} a_i x^{ik}, \quad g = \sum_{i \ge 0} b_i x^{ik}, \quad h = \sum_{i \ge 0} c_i x^{ik}, \quad \deg a_i,b_i,c_i < k. $$
 Suppose we are also given DFTs of order $3k$ (or,
alternatively, double DFTs of order $(l_1,l_2)$ with $l_1+l_2=3k$)
of all necessary blocks $a_i$, $b_i$, $c_i$. We are to show how
one can compute $q$ via approximately $3(n/k)F(k)$ extra
operations.

Flooring makes some complication. We avoid it as following. Let
$$ u_i = \sum_{\mu+\nu=m/k+i} b_{\mu}c_{\nu}, \qquad \theta = \lfloor u_{-1} / x^k \rfloor. $$
Then
$$ \lfloor g h / x^m \rfloor = \theta + \sum_{i \ge 0} u_i x^{ik}. $$
Finally we have
$$ q = \sum_{i \ge 0} d_i x^{ik} \mod x^n, \qquad d_i = a_i \theta + \sum_{\lambda+\mu=i} a_{\lambda}u_{\mu}. $$
Note that $d_i$ are the polynomials of degree $<3k$.

Turn to calculations. Let $a^*$ to denote the vector of DFT (or
double DFT) of polynomial $a(x)$.

$(i)$ Compute $u^*_i$ for $i=-1,\ldots,n/k-1$. It costs
$O(k)F((m+n)/k) = O((m+n)\log((m+n)/k))$ since $u_i^*$ are the
components of the convolution of vectors $(b_0^*, b_1^*, \ldots ,
b_{(m+n)/k-1}^*)$ and $(c_0^*, c_1^*, \ldots , c_{(m+n)/k-1}^*)$.
Recall that $b_i^*$ and $c_i^*$ are the elements of the space
$\mathbb C^{3k}$ of DFT-images with component-wise multiplication.

$(ii)$ Compute $u_{-1}$ and hence $\theta$ via inverse DFT. It
costs $3F(k)+O(k)$ as far as $u_{-1}^*$ is known (in fact,
$2k$-point inverse DFT suffices here).

$(iii)$ Compute $\theta^*$. it costs $3F(k)+O(k)$.

$(iv)$ Compute $d^*_i$ for $i=0,\ldots,n/k-1$. It costs
$O(k)F(n/k) =$ \linebreak $O(n\log(n/k))$ contributed by a
convolution of order $n/k$ in $\mathbb C^{3k}$ and $O(n/k)$ extra
additions and multiplications in the same space.

$(v)$ Compute all $d_i$ and hence $q$. It costs
$(n/k)(3F(k)+O(k))$ to compute $d_i$ and $O(n)$ to restore $q$.

Finally we have an upper bound
$$3(n/k+2)F(k)+O((m+n)\log((m+n)/k))$$
for the total complexity.

\end{document}


\title{Замечание о быстром вычислении экспоненты степенного ряда}
\date{}
\author{И. С. Сергеев}

\maketitle
\begin{abstract}
Показано, что первые $n$ членов экспоненты комплексного степенного
ряда могут быть вычислены за $(23/12+o(1))M(n)$ бинарных
арифметических операций над $\mathbb C$, где $M(n)$ обозначает
(приведенную) сложность умножения многочленов степени $<n$ при
помощи ДПФ. Также показано, что возведение ряда в постоянную
степень может быть реализовано со сложностью $(27/8+o(1))M(n)$.
\end{abstract}

\section{Введение}

Хорошо известно, что сложность вычисления многих элементарных
операций со степенными рядами, в том числе экспоненты, по порядку
такая же, как сложность умножения, см., например,~\cite{rbr,rgg}.
(Под сложностью понимается число элементов в схеме из
функциональных элементов или неветвящейся программе над
арифметическим базисом $\{ \pm, * \} \cup \{ ax \mid a \in \mathbb
C \}$. Вместо поля комплексных чисел можно рассматривать поле
действительных чисел, произвольное алгебраически замкнутое поле
или любое другое поле, допускающее ДПФ подходящего порядка.)

Предшествующие работы следуют соглашению измерять сложность
базовых операций над степенными рядами (включая экспоненту) числом
умножений эквивалентного размера. В последних работах (после 2000
г.) используются специальные умножения, основанные на ДПФ.

Функцию сложности $M(n)$ ДПФ-метода умножения можно ввести
следующим образом. Обозначим через $F^{\circ}(n)$ сложность ДПФ
порядка $n$ над $\mathbb C$, а через $F(n)$~--- приведенную
сложность, $F(n) = \min_{m \ge n} F^{\circ}(m)$. Предположим для
удобства, что $F(n)=\Omega(n\log n)$ (на самом деле, достаточно
было бы условия $F(n)=\omega(n)$) и $kF(n)=(1+o(1))F(kn)$ при $1
\le k \le \log n$. (Указанные предположения используются только
для удобства записи полученных оценок сложности и не затрагивают
другие аспекты доказательства. Запись $g=\omega(f)$ означает
$f=o(g)$.) Положим, $M(n) = 6F(n)$. В любом случае $M(n)$ служит
асимптотической верхней оценкой сложности умножения многочленов
степени $<n$.

Запишем верхнюю оценку сложности вычисления экспоненты
комплексного степенного ряда по модулю $x^n$ в виде $(A+o(1))M(n)$
и перечислим известные ранее результаты: $A=17/6$~\cite{rb2},
$A=14/3$~\cite{rvdh,rvdh2}, $A=13/6$~\cite{riss,ris10}
и~\cite{rh}. Ниже будет показано, что оценка справедлива при
$A=23/12$. Предлагаемый метод получается непосредственным
комбинированием методов~\cite{riss,ris10} и~\cite{rh}. Более
точно, схема вычислений следует алгоритму~\cite{rh}, но при этом
используются технические приемы из работ~\cite{riss,ris10}.

Также будет показано, что возведение комплексного степенного ряда
в постоянную степень по модулю $x^n$ можно реализовать со
сложностью $(B+o(1))M(n)$, где $B=27/8$. Это чуть лучше ранее
полученной оценки $B=41/12$~\cite{riss,ris10}.

Введем необходимые обозначения. Пусть $f \in \mathbb C[[x]]$.
Тогда через $f_{..n}$ будем обозначать $f \bmod x^n$, а через
$\lfloor f/x^n \rfloor$ соответственно $(f-f_{..n})/x^n$. Если $f
= \sum_{i \ge 0} f_i x^i$, то $\Delta f$, ${\bf J} f$, $\ln f$
(при $f_0=1$) и $e^f$ (при $f_0=0$) обозначают формальные
производную, интеграл, логарифм и экспоненту соответственно:
$$ \Delta f = \sum_{i \ge 1} if_i x^{i-1}, \quad {\bf J} f = \sum_{i \ge 1} \frac{f_{i-1}}i x^{i},
\quad  \ln f = - \sum_{i \ge 1} \frac{(1-f)^i}{i},  \quad e^f =
\sum_{i \ge 0} \frac{f^i}{i!}. $$

\section{Экспонента}

Рассмотрим задачу вычисления экспоненты степенного ряда $h$,
$h_{..1}=0$. Обозначим $f=e^h$, $r=1/f$. Заметим, что $\Delta
h=\Delta f/f$.

Следующая итерационная формула~\cite{rvdh} (получаемая из
уравнения вида $\Delta f = g f$, где в нашем случае $g=\Delta h$)
справедлива при $m\ge n$:
\begin{equation}
f_{..m+n} = f_{..m} + f_{..n} \, {\bf J} \left( x^{m-1}r_{..n}
\left\lfloor \Delta (h_{..m+n}) f_{..m} / x^{m-1} \right\rfloor
\right) \mod x^{m+n}.
\end{equation}

Обозначим через $E(n)$ и $I(n)$ сложность вычисления $e^f$ и $1/f$
по модулю $x^n$ соответственно. Тогда, используя~(1), можно
вычислить $f_{..m}$ со сложностью
\begin{equation}
E(n) + I(n) + (13+o(1))F(m) \sim (13/6+o(1))M(m),
\end{equation}
если параметры $m$ и $n$ выбрать подходящим образом, например,
$n=o(m)$ и $m=O(n\sqrt{\log n})$. (Это один из способов вычислить
экспоненту со сложностью $13/6$ умножений~\cite{ris10}.)

Для доказательства оценки~(2) разобьем ряды на блоки подходящего
размера $k$, например, $k \in o(n) \cap \Omega(n/\sqrt{\log n})$.
Для выполнения блочных умножений в~(1) будем использовать двойные
ДПФ порядка $(2k,k)$.

{\it Двойное ДПФ} порядка $(l,k)$ как отображение из $\mathbb
C[x]$ в $\mathbb C^{l+k}$ определяется следующим образом: первые
$l$ компонент являются компонентами ДПФ порядка $l$, другие $k$
компонент получаются композицией замены переменной $x \to \zeta x$
и ДПФ порядка $k$, где $\zeta$~--- подходящее комплексное число.
Подобным образом определяются ДПФ большей кратности. Многократные
ДПФ выгодно использовать в случае, когда выполняется несколько
умножений разной размерности, но с общими входами. Двойное ДПФ
порядка $(l,k)$, равно как и обратное к нему, реализуется со
сложностью ДПФ порядка $l$, ДПФ порядка $k$ плюс $O(l+k)$
дополнительных операций, подробнее см. в~\cite{ris10} (о
многократных ДПФ см.~\cite{ris9}).

В случае $l=2k$ (т.е. в нашем случае) вместо двойных ДПФ можно
использовать обычные ДПФ порядка $3k$, представляемые композицией
внешних ДПФ порядка $3$ и внутренних ДПФ порядка $k$.

Главный член в оценке~(2) складывается из следующих составляющих:
$3(m/k)F(k)$ (что по предположению приблизительно равно $3F(m)$)
операций для вычисления ДПФ блоков ряда $f$, столько же~--- для
вычисления ДПФ блоков ряда $\Delta h$, столько же~--- для
восстановления блоков тройного произведения под знаком интеграла,
$2F(m)$ операций для вычисления ДПФ порядка $2k$ блоков ряда ${\bf
J}(\ldots)$ и столько же~--- для реконструкции произведения
$f_{..n} \, {\bf J} (\ldots)$. Остальные действия (предвычисление
$f_{..n}$ и $r_{..n}$, реализация операторов $\Delta$ и $\bf J$,
вычисления в пространстве Фурье-образов) вносят вклад $o(m\log m)$
в общую оценку сложности.

С целью предоставить некоторое обоснование указанным оценкам в
Приложении подробнее рассматривается подзадача тройного умножения
(вероятно, наименее очевидная часть описанного алгоритма). Другие
детали доказательства могут быть восстановлены по
работе~\cite{ris10}.

Теперь изложим улучшенную версию алгоритма, используя идею из
работы Д.~Харви~\cite{rh}.

Пусть требуется вычислить $f_{..2m}$. Для этого сначала мы
вычисляем $f_{..m}$ способом, указанным выше. При этом попутно
также вычисляются ДПФ блоков $f_{..m-n}$ и $\Delta (h_{..m})$. На
втором этапе используется формула~\cite{rbr}
\begin{equation}
f_{..2m} = f_{..m} + f_{..m}(h - \ln f_{..m})_{..2m} \mod x^{2m},
\end{equation}
получаемая дискретным аналогом метода Ньютона---Рафсона из
уравнения $\varphi[f]=h$, где в нашем случае $\varphi[x]= \ln x$.
Этап состоит из двух основных частей: вычисление $\ln f_{..m}$ по
модулю $x^{2m}$ и заключительное умножение $f_{..m}$ на $h-\ln
f_{..m}$.

Обозначим $s=\Delta(f_{..m})/f_{..m}$. Тогда $(\ln
f_{..m})_{..2m}$ вычисляется как ${\bf J} s_{..2m-1}$ при помощи
итерационной формулы~\cite{rkm}
\begin{equation}
 s_{..m'+n-1} = s_{..m'-1} - x^{m'-1}r_{..n}\left\lfloor s_{..m'-1}f_{..m} / x^{m'-1} \right\rfloor \mod x^{m'+n-1},
\end{equation}
где $m'\ge m$. Начало ряда $s$ известно: $s_{..m-1} = \Delta
(h_{..m})$.

Для вычислений по формуле~(4) используется дополнительно
$(6+o(1))F(m)$ операций: половина из них~--- для вычисления
недостающих ДПФ блоков ряда $s$, другая половина~--- для
восстановления блоков тройного произведения (о тройном умножении
см. в Приложении).

Для завершения вычисления $f_{..2m}$ выполняется еще
$(4+o(1))F(m)$ операций: половина из них~--- для вычисления ДПФ
порядка $2k$ блоков ряда $h-\ln f_{..m}$, примерно столько же~---
для восстановления блоков произведения в формуле~(3). Заметим, что
ДПФ порядка $2k$ почти всех блоков $f_{..m}$ вычислены на первом
этапе алгоритма как части двойных ДПФ (это то место, где
применение двойных ДПФ дает выигрыш).

Окончательно для вычисления первых $2m$ членов ряда экспоненты мы
получаем оценку сложности $(23+o(1))F(m)$.

\section{Возведение в степень}

Рассмотрим задачу возведения ряда $h$, $h_{..1}=1$ в степень $C
\in \mathbb C$. Обозначим $f=h^C$, $r=1/f$, $\rho = 1/h$,
$s=C\Delta h/h$.

Для возведения в степень можно предложить алгоритм, аналогичный
описанному выше алгоритму вычисления экспоненты. Изложим его
кратко.

Младшая половина искомых коэфициентов ряда $f$ вычисляется по
формуле
\begin{equation}
f_{..m+n} = f_{..m} + f_{..n} \, {\bf J} \left( x^{m-1} r_{..n}
\left\lfloor s_{..m+n-1}f_{..m} / x^{m-1} \right\rfloor \right)
\mod x^{m+n},
\end{equation}
основанной на соотношении $\Delta f = s f$. Дальнейшие вычисления
выполняются по формуле
\begin{equation}
f_{..2m} = f_{..m} + f_{..m}({\bf J} s - \ln f_{..m})_{..2m} \mod
x^{2m},
\end{equation}
которая выводится из уравнения $\ln f = C\ln h$.

Для вычисления ряда $s$ используется итерационная
формула~\cite{rkm}
\begin{equation}
 s_{..m+n-1} = s_{..m-1} + \rho_{..n}(\Delta(h_{..m+n}) - s_{..m-1}h_{..m+n}) \bmod x^{m+n}.
\end{equation}

Как и прежде, все ряды разбиваются на блоки длины $k$, за
исключением первой половины ряда $\Delta h$, которая разбивается
на блоки длины $2k$. Затем в вычислениях используются двойные ДПФ
порядка $(2k,k)$.

Пусть $f_{..n}$, $r_{..n}$, $\rho_{..n}$ и $s_{..n-1}$ даны. Тогда
мы можем вычислить $f_{..2m}$ за $(40,5+o(1))F(m)$ дополнительных
операций. Эта оценка складывается из следующих слагаемых:

$(10,5+o(1))F(m)$ операций для вычисления $s_{..m-1}$ и ДПФ блоков
$s_{..m-1}$ при помощи~(7). Для этого вычисляются ДПФ блоков рядов
$h$, $s$, $\Delta h$, $\rho$ и обратные ДПФ для реконструкции
тройных произведений (оценка достигается благодаря тому, что блоки
ряда $\Delta(h_{..m})$ имеют двойной размер);

$(10+o(1))F(m)$ операций для вычисления старшей половины
$s_{..2m-1}$. Здесь каждая итерация по формуле~(7) выполняется за
два обычных умножения, для чего используются ДПФ порядка $2k$;

$(10+o(1))F(m)$ операций для вычисления частичного ряда $f_{..m}$
и ДПФ его блоков~(5). Процедура такая же, как в первой части
алгоритма вычисления экспоненты;

$(6+o(1))F(m)$ операций для вычисления $2m$ первых членов ряда
$\ln f_{..m}$. Этот шаг совпадает с соответствующим шагом
описанного выше алгоритма для эспоненты;

$(4+o(1))F(m)$ операций для реализации заключительного умножения
по формуле~(6) при помощи ДПФ порядка $2k$.

Суммируя, получаем заявленную оценку.

Работа выполнена при финансовой поддержке РФФИ, проекты
11--01--00508 и 11--01--00792--а, и программы фундаментальных
исследований ОМН РАН <<Алгебраические и комбинаторные методы
математической кибернетики и информационные системы нового
поколения>> (проект <<Задачи оптимального синтеза управляющих
систем>>).

\section*{Приложение}

Пусть $f,g,h \in \mathbb C[[x]]$, и требуется вычислить
$$ q = f \lfloor g h / x^m \rfloor \mod x^n, $$
где $n|m$. Пусть ряды $f,g,h \in \mathbb C[[x]]$ разбиты на блоки
длины $k$ (для простоты мы полагаем, что $n$ делится на $k$):
$$  f = \sum_{i \ge 0} a_i x^{ik}, \quad g = \sum_{i \ge 0} b_i x^{ik}, \quad h = \sum_{i \ge 0} c_i x^{ik}, \quad \deg a_i,b_i,c_i < k. $$
 Пусть также даны вектора ДПФ порядка $3k$
(или двойных ДПФ порядка $(l_1,l_2)$, где $l_1+l_2=3k$) всех
необходимых блоков $a_i$, $b_i$, $c_i$. Покажем, что $q$ можно
вычислить приблизительно за $3(n/k)F(k)$ дополнительных операций.

Некоторые трудности создает отбрасывание младших членов ряда $gh$.
Их можно преодолеть следующим образом. Обозначим
$$ u_i = \sum_{\mu+\nu=m/k+i} b_{\mu}c_{\nu}, \qquad \theta = \lfloor u_{-1} / x^k \rfloor. $$
Тогда
$$ \lfloor g h / x^m \rfloor = \theta + \sum_{i \ge 0} u_i x^{ik}. $$
Окончательно получаем
$$ q = \sum_{i \ge 0} d_i x^{ik} \mod x^n, \qquad d_i = a_i \theta + \sum_{\lambda+\mu=i} a_{\lambda}u_{\mu}. $$
Заметим, что $d_i$~--- многочлены степени $<3k$.

Опишем порядок вычислений. Через $a^*$ будем обозначать вектор ДПФ
(или двойного ДПФ) многочлена $a(x)$.

$(i)$ Вычисляются $u^*_i$, где $i=-1,\ldots,n/k-1$. Сложность
можно оценить как $O(k)F((m+n)/k) = O((m+n)\log((m+n)/k))$, т.к.
$u_i^*$~--- компоненты свертки векторов $(b_0^*, b_1^*, \ldots ,
b_{(m+n)/k-1}^*)$ и $(c_0^*, c_1^*, \ldots , c_{(m+n)/k-1}^*)$.
Напомним, что $b_i^*$ и $c_i^*$ принадлежат пространству
Фурье-образов $\mathbb C^{3k}$ с покомпонентной операцией
умножения.

$(ii)$ Вычисляется $u_{-1}$ и, следовательно, $\theta$ при помощи
обратного ДПФ. Сложность оценивается как $3F(k)+O(k)$, т.к. вектор
$u_{-1}^*$ известен (в действительности, достаточно выполнить
обратное ДПФ порядка $2k$).

$(iii)$ Вычисляется $\theta^*$. Сложность $3F(k)+O(k)$.

$(iv)$ Вычисляются $d^*_i$, где $i=0,\ldots,n/k-1$. Сложность
$O(k)F(n/k) = O(n\log(n/k))$ складывается из сложности свертки
порядка $n/k$ в $\mathbb C^{3k}$ и $O(n/k)$ дополнительных
сложений в том же пространстве.

$(v)$ Вычисляются все $d_i$ и затем $q$. Здесь $(n/k)(3F(k)+O(k))$
операций используется для вычисления $d_i$ и $O(n)$~--- для
восстановления $q$.

Окончательно для алгоритма в целом получаем оценку сложности
$$3(n/k+2)F(k)+O((m+n)\log((m+n)/k)).$$